# Students' understanding of the 2D Heat Equation: An APOS approach

Maria Al Dehaybes[1], Johan Deprez[2], Paul van Kampen[3], and Mieke De Cock[1]

[1] *LESEC & Department of Physics and Astronomy, Faculty of Science, KU Leuven, 3001 Leuven, Belgium*
[2] *LESEC & Department of Mathematics, Faculty of Science, KU Leuven, 3001 Leuven, Belgium*
[3] *CASTeL & School of Physical Sciences, Dublin City University, Ireland*

In this paper, we use the APOS theoretical framework to validate a hypothetical learning trajectory of the 2D heat equation, a preliminary genetic decomposition that stresses the conceptual understanding of its mathematical formulation. We design questions to probe specific mental constructions of the preliminary genetic decomposition. We interview 8 students in the second year of their B.Sc. enrolled in either engineering, physics, or twin (mathematics and physics) majors. Our findings indicate that students engage with many predicted mental constructions. In particular, coordination and encapsulation of two process conceptions of the Laplacian of the temperature improve understanding although it is challenging. Other parts of the genetic decomposition require refinements. These include mental constructions related to the temperature distribution function, heat flow, and the temperature gradient .

## I. INTRODUCTION

Combining mathematics and physics is an important but challenging aspect of learning physics for students at different education levels [1-11]. In this project, we investigate students' understanding of the 2D heat equation, a context in which it is important to combine elements from the two disciplines.

For this purpose, we use a theoretical framework from mathematics education research: APOS. We chose this theoretical framework to employ one of its most important tools, the genetic decomposition (Fig. 1). The genetic decomposition is a model of the mental constructions that a student needs to make to learn a certain concept. A novel aspect in our approach is the extension of the genetic decomposition to include mental constructions requiring the combination of mathematics and physics. We have used the 2D heat equation as a first application of this extended APOS framework.

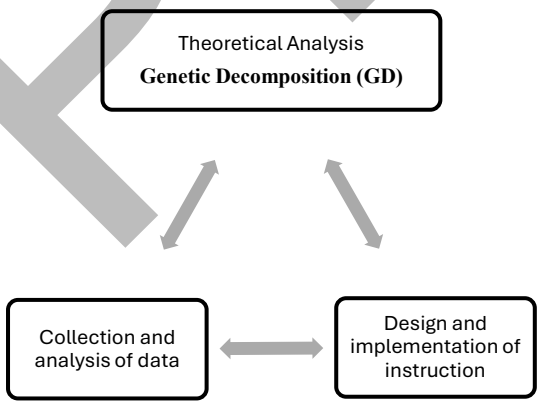

FIG. 1. Research cycle in APOS [12].

APOS is a theoretical framework that combines three components (see Fig. 1): (1) a theoretical analysis, (2) data gathering, and (3) the design of instruction. It starts with the theoretical analysis of a concept through the development of a preliminary genetic decomposition (PGD). The preliminary genetic decomposition includes mental constructions which serve as the base for a hypothetical learning trajectory of a concept. This is followed by the collection and analysis of data that could lead to an alteration of the theoretical analysis. The reviewed theoretical analysis is used to design instruction, which is to be implemented in class. New data can be collected from this implementation. This could in turn again lead to alterations in the theoretical analysis. These three components allow APOS to combine the investigation of students' reasoning and the development of topic-based research-validated instructional materials.

We have described in an earlier paper [13] how we extended APOS to include physics concepts and physico-mathematical objects. In that paper we proposed a possible learning trajectory for the heat equation in 2D. In this paper, we focus on the validation of this hypothetical learning trajectory. Through carefully designed tasks, we prompted students via task-based think aloud interviews to carry out specific mental constructions. The results of these interviews have informed revisions of the preliminary theoretical analysis, a common step in the development of teaching and learning materials and strategies in APOS theory.

Briefly, the 2D heat equation,

$$\frac{\partial T}{\partial t} = \alpha \nabla^2 T, \quad (1)$$

describes how the temperature distribution on a 2D plate evolves over time due to heat conduction. It relates the temporal rate of change of the temperature, $\frac{\partial T}{\partial t}$, at a specific point on a 2D plate with a given temperature distribution, to the Laplacian of the temperature at that particular point, $\nabla^2 T$ (also denoted by $\Delta T$, which is the form used in the courses taken by our students, or $\vec{\nabla} \cdot \vec{\nabla} T$).

Mathematically, the Laplacian is a differential operator given by the divergence of the gradient of a scalar function in Euclidean space. In a Cartesian coordinate system, the Laplacian can be calculated as the sum of the second partial derivatives of the function with respect to each independent variable. Given a scalar temperature distribution, the Laplacian of the temperature is related to the net heat flow at a point, and to the difference between the temperature at a point and the average temperature of its surroundings. Reasoning with these different interpretations of the Laplacian of the temperature in the context of the heat equation involves different external representations: graphical representations (3D graphs, gradient vector fields, Fig. 2), linguistic descriptions, and symbolic representations.

For example, graphical and linguistic representations could be combined as follows. On a Cartesian graph of the temperature distribution (Fig. 2(a)), the temperature at a point can be compared to the average temperature of its surroundings. Heat will flow into a point from regions of higher temperature and away from the same point to regions of lower temperature in accordance with Fourier's law of heat conduction. Since the temperature at point $J$ in Fig. 2(a) is lower than the average temperature of the points surrounding it, heat will flow in, and the *net* heat flow at this point is positive. This causes an increase in the temperature at $J$ as time passes, and according to the heat equation, the Laplacian at $J$ is positive.

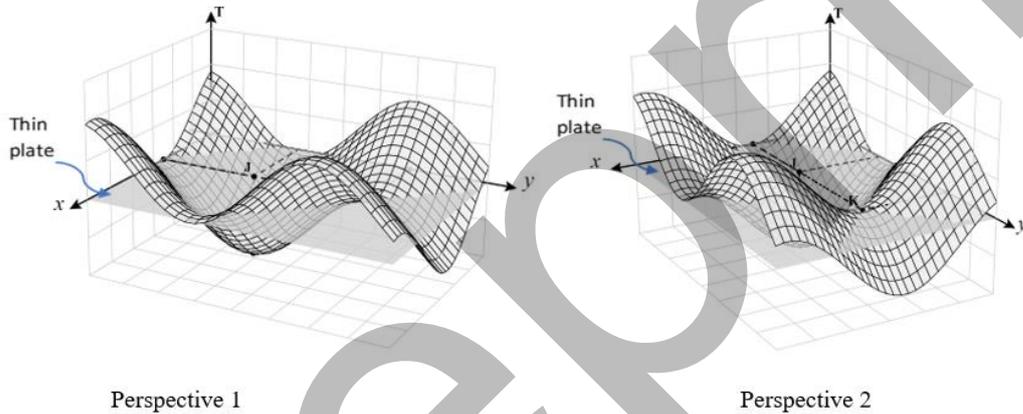

(a)

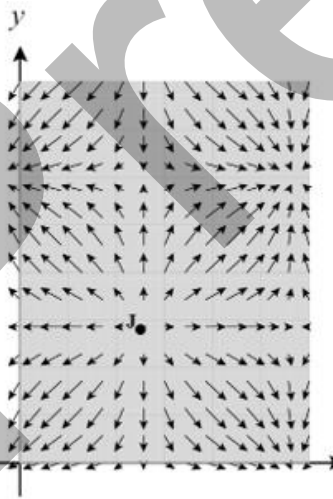

(b)

FIG. 2. Temperature distribution on a 2D plate at time $t = t_1$ represented (a) in a Cartesian graph and (b) in a gradient vector field plot

Symbolical and graphical representations could be combined as follows. The Laplacian of the temperature at a point on a 2D plate can be expressed as:

$$\nabla^2 T = \frac{\partial^2 T}{\partial x^2} + \frac{\partial^2 T}{\partial y^2}. \quad (2)$$

Moving in the direction of the positive $x$-axis or $y$-axis around $J$ in Fig. 2 and comparing slopes of tangents, the spatial rate of change of the temperature is increasing (since it becomes less negative) as one approaches $J$. The Laplacian, being the sum of second order PDs, is therefore positive, and, according to equation (1), the temperature at point $J$ will increase with respect to time. At other points, like $K$, what happens is less obvious. There are regions from which heat is flowing towards $K$ and regions to which heat is flowing away from $K$. Therefore, the net heat flow at $K$ could be positive, negative, or zero, depending on the "average bending" along $x$ and $y$ at point $K$.

To link temperature change and heat flow, students need to combine Fourier's law of heat conduction and the definition of specific heat capacity. According to Fourier's law of heat conduction, the temperature gradient $\vec{\nabla} T$ is proportional to the heat flux density, $\vec{q}$:

$$\vec{q} = -k\vec{\nabla} T \quad (3)$$

where $k$ is the thermal conductivity of the material. The divergence of the heat flux gives the net heat per unit volume flowing out of a point. The temperature increase $\Delta T$ of an object of mass $m$ and specific heat capacity $c$ is related to the net heat $Q$ flowing into it by $Q = mc\Delta T$. Therefore, the temporal rate of change of the temperature is related to the divergence of the heat flux density by

$$\frac{\partial T}{\partial t} = -\frac{\vec{\nabla} \cdot \vec{q}}{\rho c} = \frac{k}{\rho c}\vec{\nabla} \cdot \vec{\nabla} T = \alpha \vec{\nabla} \cdot \vec{\nabla} T \quad (4)$$

where $\rho$ is the density of the material. For this reason, it is productive to think of the Laplacian as the divergence of the gradient, in this case

$$\nabla^2 T = \vec{\nabla} \cdot (\vec{\nabla} T). \quad (5)$$

or this reason, a temperature gradient vector field plot (Fig. 2b) is another useful representation of the temperature distribution on a plate. It allows for the visual identification of both the heat flow (in the opposite direction of the direction of the vectors) and the divergence of the gradient, i.e. how the temperature at each point changes. There are two common approaches to interpreting vector field plots. Following a differential approach, the divergence of the gradient of the temperature can be obtained by tracking how $\frac{\partial T}{\partial x}$ and $\frac{\partial T}{\partial y}$, the first spatial partial derivatives of the temperature, change with respect to each spatial direction, which is in accordance with equation (2). Fig. 2(b) shows that both $\frac{\partial T}{\partial x}$ and $\frac{\partial T}{\partial y}$ increase in the direction of $+x$ and $+y$ near point $J$. Following an integral approach, the Laplacian of the temperature at a point can be obtained by analyzing the flux of the temperature gradient through a closed surface surrounding this point. The flux at point $J$ is positive, and therefore the divergence of the temperature gradient and the Laplacian are, too.

TABLE I. Concepts related to the understanding of the 2D heat equation.

| Concept | Importance in the 2D heat equation |
|---|---|
| First partial derivative | Spatial and temporal rate of change of temperature. |
| Laplacian of temperature | How heat diffuses in 2D space, net heat flow at a point. |
| Spatial second partial derivative | Concavity of temperature distribution at a point affects heat diffusion rate. |
| Divergence | Divergence of heat flux density, divergence of $T$ gradient. Derivation of the heat equation: Divergence theorem relates behavior of heat flux inside a volume to net heat flow at boundaries. |
| Temperature gradient | Steepest spatial rate of change of temperature at a point. Perpendicular to isotherms. Has a direction opposing heat flux density. |
| Fourier's law | Links temperature gradient and heat flux density. |
| Multivariable functions and 3D space | Representation of $T$ distribution. |

In this work, we focus on the conceptual understanding of the mathematical formulation of the heat equation and how it relates to the underlying physics concepts. We are not concerned with the solution procedure. In Table I, we include a list of mathematical and physical concepts relevant to the development of this understanding. As outlined above, students need a physical understanding of temperature distribution, heat transfer, and Fourier's law. They also need a mathematical understanding of concepts such as the first and second partial derivatives, gradient, divergence, and Laplacian, including their linguistic descriptions and graphical representations.

In Section II, we describe the mental constructions of the theoretical framework, discuss our extension of the framework into the context of the 2D heat equation, and outline our preliminary genetic decomposition. In Section III, we describe how the design of the questions to validate our preliminary genetic decomposition stems from this preliminary genetic decomposition, and the administration of questions in interviews. In Section IV, we report on findings question by question, and how these findings can be used to refine the genetic decomposition. In Section V, we conclude our findings.

## II. THEORETICAL FRAMEWORK

### A. Action, Process, Object, Schema (APOS)

In our study, we use APOS, a theoretical framework from mathematics education research. APOS describes how individuals learn (mathematical) concepts through mental constructions called structures and mechanisms. Mental structures are stable and are used to make sense of a situation, while mechanisms are tools through which a structure is developed. Mental structures include action, process, object, and schema. Mental mechanisms include interiorization, encapsulation, de-encapsulation, coordination, and thematization. A set of hypothesized mental constructions needed to learn a concept is called a preliminary genetic decomposition (PGD). We will explain those mental constructions relevant to our design using fractions as an example. Students often learn about fractions in terms of cutting pizza.

In an **action conception** (see Fig. 3), a learner performs actions based on external instructions. For instance, using a pizza and fractions example, this means physically cutting the pizza into equal parts while observing the slices. In a **process conception**, a learner reflects on their actions and identifies patterns. They understand that cutting a pizza into $n$ equal parts means that each slice represents a fraction $1/n$ of the whole. In this way a learner has interiorized the action into a process. In an **object conception**, a learner encapsulates the process into a mental object which means they have transformed the dynamic process into a static object. Fractions like 1/2, 1/3, and 1/4 are understood as static mathematical entities that summarize symbolically the dynamic process of dividing a whole into parts. The learner can perform operations on these objects, such as adding fractions. The learner can coordinate more than one process and encapsulate it in an object conception. The learner can de-encapsulate the object back into a process or processes. A **schema conception** integrates all these understandings into a cohesive mental framework. Extending this to the example of fractions, this framework connects fractions as parts of a whole, equivalent fractions, operations on fractions, and their representation as static structures. It allows for flexible reasoning and application of fractions across different contexts. A learner can thematize a schema into an object by performing actions on it. In the case of fractions, students can represent a fraction using visual representations, e.g., a fraction bar or pie chart.

TABLE II. APOS definitions.

| APOS Mental Structures | APOS Definitions and keywords |
| --- | --- |
| Action conception | Procedural understanding, step by step, follow instructions. |
| Process conception | Interiorized action, de-encapsulated object. |
| Object conception | Encapsulated process, thematized a schema, coordinated processes. |
| Schema conception | Coherent collection of processes, objects, actions and relative mechanisms. Connected concepts. Coherence could possibly lead to a decision whether a problem falls within schema |

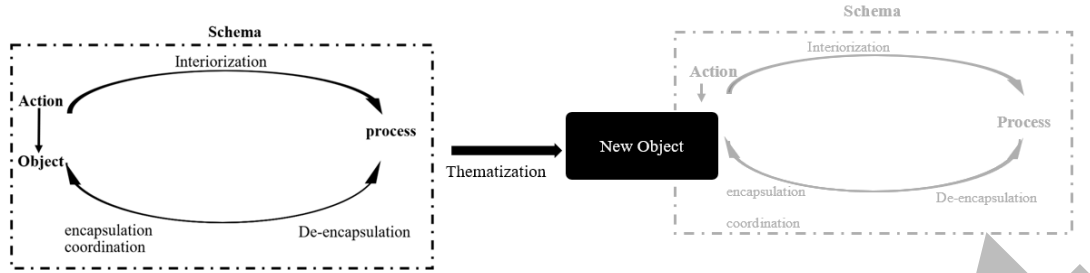

FIG. 3 Part of mental structures and mechanisms and how they are related [14]

### B. Extending APOS to describe the interplay of mathematics and physics

The APOS theoretical framework builds on Piaget's concept of reflective abstraction [14]. In a previous paper [13], we described in detail how we extended APOS into physics education. Here we focus on one aspect of APOS theory: a researcher can hypothesize a preliminary genetic decomposition, test it by collecting and analyzing data, and review it. Research instruments to test the preliminary genetic decomposition are specifically designed to probe the hypothesized mental constructions of learners. The predicted mental constructions are validated if learners actually construct them. If students face difficulties, or show evidence of making different mental constructions, the preliminary genetic decomposition may be altered; it is then called a reviewed genetic decomposition. Learning and teaching materials are then redesigned and tested in class according to the revised genetic decomposition. The cycle is repeated until a stable model is achieved (see Fig. 4).

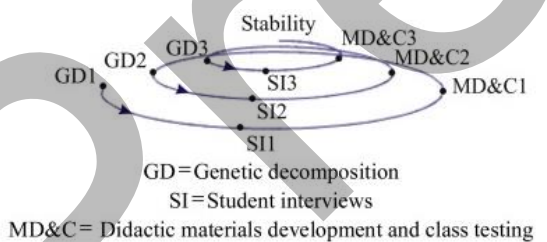

FIG. 4. Cycle of research in APOS [15].

In our extension of APOS to the teaching and learning of the 2D heat equation in physics, we have embedded the integration of mathematics and physics not only at the schema conception, but also at the object and process conceptions. We hypothesized that embedding the integration of both disciplines throughout would lead to a deep understanding by the students. We know from the literature that even students who show understanding of a concept in a mathematical context do not automatically show understanding when the concept is applied in a physics context [4, 16-18]. Likewise, using mathematics as purely a tool for physics is also often ineffective [19].

When designing the preliminary genetic decomposition, we considered four aspects: (1) Concepts from mathematics and physics necessary to develop a conceptual understanding of the 2D heat equation statement, (2) previously known student challenges related to these concepts, (3) multiple external representations of concepts, and (4) extension of APOS to a new context where the interplay of mathematics and physics is crucial. In this paper, we elaborate on the content of the preliminary genetic decomposition necessary for the design of interview questions (mainly aspects (1) and (2)). While aspects (3) and (4) were indeed essential—particularly in formulating interview questions that probe students' mental constructions at different APOS conceptions—a separate paper is dedicated to discussing this theoretical extension in depth [13]. There, we elaborate on the rationale for the decisions related to multiple external representations when designing the questions, the interplay of mathematics and physics through the lens of APOS, and the operationalization of APOS to describe students' understanding of mathematics and physics in the context of the 2D heat equation. We also discuss the extension of APOS to physico-mathematical processes, objects, and schemata relevant to the 2D heat equation in the same paper.

### C. Preliminary genetic decomposition of the 2D heat equation

In our preliminary genetic decomposition of the 2D heat equation, we differentiated between concepts that are needed to understand the statement of the 2D heat equation, and prerequisite concepts. Concepts directly tied to the structure of the two-dimensional heat equation—such as the temperature function, its temporal and spatial rates of change, the gradient of the temperature, the divergence, and the Laplacian of the temperature—are treated as core components of the 2D heat equation. In contrast, we regard other concepts typically introduced in prior courses, such as the second partial derivative, as prerequisites. This distinction is best understood as flexible. It is adopted here to support analytical clarity rather than to imply fixed or mutually exclusive categories. We developed the preliminary genetic decomposition through a detailed analysis of the core concepts involved in the 2D heat equation, combined with insights about students' difficulties from previous studies [10,16,20-25], as well as related research on the use of multiple external representations to aid in conceptual understanding [7,26-30].

We first describe the concepts directly tied to the structure of the two-dimensional heat equation with reference to specific mental constructions which are included in Table III(a).

Students need to understand the heat equation as a relationship between the Laplacian of the temperature $\nabla^2 T$ and the temporal rate of change of the temperature, $\frac{\partial T}{\partial t}$ (HE10). They need to know how to interpret the Laplacian of the temperature as related to the net heat flow at a point and to the temperature difference between a point and the average of its surroundings (HE9). In a previous study [31], we asked students how they interpret the Laplacian of the temperature. Many used the word "rate" to describe the Laplacian, suggesting that they interpret it as similar or identical to other concepts such as the (partial) derivative or the gradient. In addition, many could not link the Laplacian graphically to concavity, or to the temperature difference between a point and the average of its surroundings. To develop this understanding, we hypothesize that students need to understand the Laplacian of the temperature as the "average bending" along the two spatial variables $x$ and $y$ on a 3D graph (HE8). They therefore need to reason with the Laplacian as the sum of second partial derivatives (HE7 and HE6). We also found that almost no students linked the Laplacian to the net heat flow. We hypothesize that students need to reason with the Laplacian of the temperature as the divergence of the temperature gradient (HE5), so they may relate it to the net heat flow and the heat flux density.

To develop an understanding of the divergence of the temperature gradient, students need to understand the temperature gradient itself (HE4). In a previous study [31], we asked students to interpret the temperature gradient. Similar to the Laplacian, students used the word "rate" to describe it. They often did not precisely articulate their understanding of the gradient as the steepest ascent at a point. Therefore, we include HE4 in our genetic decomposition. Of course, to reason with all that we mentioned, students need an understanding of temporal rate of change (HE3), spatial rate of change (HE2), and temperature distribution as a multivariable function (HE1). We know from previous literature [21] that students interpret heat flow as a time derivative of temperature instead of a spatial derivative. Therefore, we include both the temporal and the spatial rate of change in our genetic decomposition as HE2 and HE3.

In what follows, we describe the second category, the prerequisite concepts shown in Table III(b). To develop an understanding of the physical meaning of the heat equation, students need to understand heat conduction (Pre-i). To develop an understanding of the Laplacian of the temperature, the divergence, and the temperature gradient, they need to know the symbolic representation of these concepts and how to use different variables (temperature, position, ...) within their symbolic representation (Pre-ii). To develop an understanding of the temporal rate of change of the temperature and the spatial rate of change of the temperature, students need an understanding of first partial derivates as calculations of rate of change and as slopes of tangents at a point (Pre-iii). One possible way to develop an understanding of the Laplacian of the temperature as "average bending" is to treat the Laplacian as the sum of second partial derivatives. This path is only accessible to students if they understand second partial derivatives (Pre-iv). Finally, to develop an understanding of temperature as a multivariable function, students need an understanding of symbolic and graphical representations of functions in 3D (Pre-v and Pre-vi).

Given that it is essential to explore how students cognitively navigate the mental constructions hypothesized in our preliminary genetic decomposition, we formulate the following research questions:

RQ1. How do students engage with the mental constructions in the preliminary genetic decomposition of the 2D heat equation?

RQ2. What key insights from the students' engagement with the preliminary genetic decomposition can inform the refinement of the genetic decomposition and the development of effective learning material?

Table III.(a) Preliminary genetic decomposition of the 2D Heat Equation

| | **Heat Equation Mental Constructions** |
|---|---|
| **HE1** | A schema understanding of temperature distribution which includes<br>   a. the multivariable function as a sub-schema<br>   b. 3D space as a sub-schema.<br>   c. an object conception of temperature. |
| **HE2** | An object conception of heat flow as a first partial derivative with respect to spatial variables $\frac{\partial T}{\partial x}, \frac{\partial T}{\partial y}$. This object conception arises from coordinating the process conceptions of:<br>   a. Fourier's law<br>   b. heat<br>   c. 1st partial derivative (Pre-iii.a) |
| **HE3** | An object conception of the temporal rate of change of temperature as $\frac{\partial T}{\partial t}$. This object conception arises from thematizing the first partial derivative schema to include time as an independent variable. (prerequisite iii) |
| **HE4** | A schema conception of the temperature gradient that includes:<br>   a. An object conception of the vector field plot representation of a gradient and its de-encapsulation into the process of obtaining the magnitude and direction of the steepest rise at a point.<br>   b. An object conception of the mathematical expression of the temperature gradient. The object conception arises from coordinating processes in HE4a and Pre-vi.d.<br>   c. A process conception of the temperature gradient as interiorizing the action of calculating the spatial rate of change of $T$ at a point.<br>   d. Coordinating the processes named in HE4a and Pre-vi.d to produce a new object: the temperature gradient as the highest spatial rate of change of temperature at a point. |
| **HE5** | A process conception of the Laplacian of the temperature as the divergence of the gradient of $T$. A learner interiorizes how $\vec{\nabla} T$ is affected by the divergence operator without performing calculations. |
| **HE6** | An object conception of spatial concavity along 1 spatial variable as how the spatial rate of change varies with respect to a spatial variable. |
| **HE7** | A process conception of the Laplacian of the temperature as the sum of second partial derivatives of $T$ with respect to independent spatial variables. |
| **HE8** | An object conception of the Laplacian of the temperature as the "average bending" along 2 independent spatial variables at a point. |
| **HE9** | Coordination of process HE5 and process HE7 to produce a new object that encapsulates the physical interpretation of the Laplacian of the temperature regardless of the representation: how the temperature at a point compares to the temperature of its surroundings. |
| **HE10** | De-encapsulation of the symbolic representation of the heat equation object into a process conception of the heat equation. |

Table IV.(b) Prerequisites of the 2D Heat Equation

| | **Prerequisites** |
|---|---|
| **Pre-i** | A schema conception of heat conduction that includes: <br> a. a process conception of Fourier's law so that a student can predict the heat flow (flux) based on the temperature gradient without the need for calculation. <br> b. a process conception of temperature as a measure of thermal energy and process conception of heat as a transfer of thermal energy; to differentiate the 2 concepts. <br> c. a process conception of thermal equilibrium to reflect on how the balance of temperature leads to no heat transfer. |
| **Pre-ii** | A schema conception for each of the following concepts: Laplacian, gradient, and divergence. Each schema includes: <br> a. a process conception to reflect on how changing independent variables and dependent variables affects them. <br> b. an object conception of Laplacian, gradient and divergence as symbolic representations. |
| **Pre-iii** | A schema conception of 1$^{st}$ partial derivative that includes: <br> a. a process conception of calculating the rate of change <br> b. an object conception of the slope of the tangent at a point. |
| **Pre-iv** | A schema conception of 2$^{nd}$ partial derivative that includes: <br> a. a process conception of obtaining the sign of the concavity along an independent variable <br> b. an object conception of concavity along one independent variable <br> c. an object conception of rate of change of rate of change as an object. |
| **Pre-v** | A schema conception of a multivariable function that includes: <br> a. an object conception of symbolic representation <br> b. an object conception of graphical representation |
| **Pre-vi** | A schema for three-dimensional space that includes: <br> a. fundamental planes as geometric objects <br> b. the thin plate as a spatial object <br> c. a process conception of intersecting fundamental planes with surfaces to describe the surface geometrically or symbolically <br> d. a process conception of assigning a symbolic variable to a spatial variable |

## III. METHODOLOGY

To test our preliminary genetic decomposition, we designed four questions to probe specific mental constructions. In the following, we discuss the design of three of these questions. The fourth question is omitted since the student responses do not contribute meaningfully to this paper. We explain which mental construction of the preliminary genetic decomposition each part addresses. We then describe the interview participants and their educational background.

### A. Design of Questions

*1. Question 1*

The main objective of this question (see Fig. 5) is to probe mental constructions from the schema of the temperature gradient (HE4) and mental constructions from the schema of heat conduction (Pre-i.a and Pre-i.c). Students were presented with Cartesian graphs of a temperature distribution on two plates, A and B. Two points, $P$ and $Q$, were indicated on the thin plates. Each part (a, b, and c) of the question was explicitly designed to probe a specific mental construction related to key ideas including Fourier's law, thermal equilibrium, vector representation, and symbolic representation of the temperature gradient. Additionally, the question as a whole may engage mental constructions beyond these focal points such as the linguistic description of the temperature gradient and the understanding of the gradient as a spatial rate.

The law of heat conduction (Fourier's law) states that $\vec{q} = -k\vec{\nabla}T$. $k$ is the thermal conductivity of the material. Figures 1 and 2 represent the temperature distribution at time $t = t_1$ on two identical and rectangular plates A and B. $P$ and $Q$ are points on the plates.

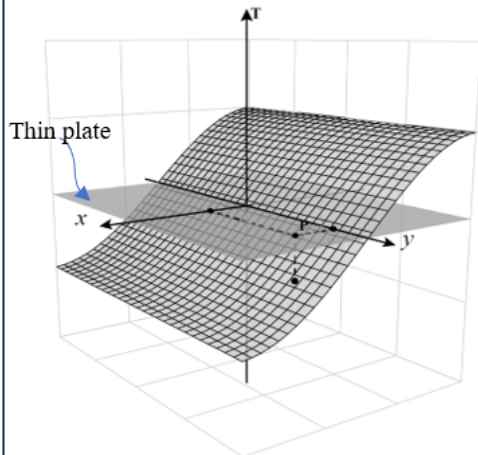

Fig1: Temperature distribution on plate A at $t = t_1$

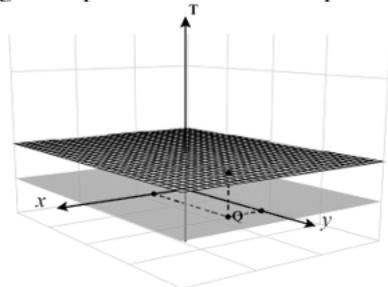

Fig 2: Temperature distribution on plate B at $t = t_1$.

a. Without calculation, explain how you can obtain the heat flux density at point $P$.
b. Without calculation, explain how you can obtain the heat flux density at point $Q$.
c. Which of the following describe the temperature gradient? Choose all that apply. Explain.

i. $\nabla T = \left(\frac{\partial T}{\partial x}; \frac{\partial T}{\partial y}\right)$   ii. $\nabla T = \left(\frac{\partial T}{\partial x}; \frac{\partial T}{\partial y}; \frac{\partial T}{\partial t}\right)$

FIG. 5. Question 1 used to probe mental constructs related to the temperature gradient hypothesized in the preliminary genetic decomposition. Students were reminded that $\vec{q}$ represents the heat flux density whenever necessary.

*2. Question 2*

The main objective of this question (see Fig. 6) was to probe mental constructions from the heat conduction schema, specifically those related to differentiation of heat and temperature at a process level (Pre-i.b). We also aimed to probe students' understanding of heat flow as proportional to the spatial rate of change of the temperature (HE2), and to see if they differentiated spatial and temporal rates of change in temperature (HE3).

Fourier's law states that the heat flux density — heat flow per unit time per cross-sectional area — is proportional to the negative temperature gradient. A two-dimensional thin rectangular plate shown in figure 1 has a temperature distribution $T$ around its boundaries so that: boundary $A$ is insulated, boundary $B$ is immersed in ice.

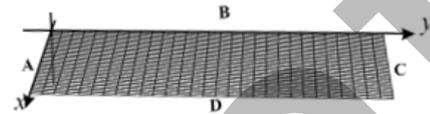

Fig 1: Thin rectangular plate

a. Which of the following is true if **boundary A** of the plate is insulated?

i. The heat passing through boundary A is zero.
ii. The thermal energy at boundary A is non-zero and constant with time.
iii. The thermal energy transfer through boundary A is zero.
iv. The temperature at boundary A is $0°C$.

b. Which of the following is true if **boundary B** of the plate is immersed in ice?

i. The heat passing through boundary B is zero.
ii. The thermal energy at boundary B is constant and non-zero.
iii. The thermal energy transfer through boundary B is zero.
iv. The temperature is $0°C$.

c. Which of the following expressions represent "**boundary A** of the plate is insulated"?

i. At boundary A $\frac{\partial T}{\partial x} = 0$
ii. At boundary A $\frac{\partial T}{\partial y} = 0$
iii. At boundary A $\frac{\partial T}{\partial t} = 0$

d. Which of the following expressions represent "**boundary B** of the plate is immersed in ice"?

i. $T(0, y, t) = 0$
ii. $T(x, 0, t) = 0$
iii. $T(x, y, 0) = 0$
iv. $T(x, y, t) = 0$

FIG. 6. Question 2 used to probe mental constructs related to heat conduction hypothesized in the preliminary genetic decomposition.

Throughout this question, we implicitly probed the schema conception of 1st PD (Pre-iii).

### 3. Question 3

In our preliminary genetic decomposition, we predicted that students need to coordinate their process understanding of the Laplacian as the divergence of the gradient with their process understanding of the rate of change of the rate of change of the temperature (HE9). In this question, we probe the mental constructions underlying this coordination. Therefore, we designed it to progressively build the foundational elements for developing this conception. Given the length of the question, we place its figure alongside its corresponding finding for clarity. We started by probing the process conception of the Laplacian of the temperature as the divergence of the gradient (HE5), see Fig. 8. Next, we probed the schema conception of the second partial derivative (Pre-iv) and spatial concavity along 1 spatial variable (HE6), including the sign of concavity, concavity as an object, and rate of change of rate change (see Fig. 9). This is followed by probing the process conception of the Laplacian of the temperature as the sum of second partial derivatives (HE7) and the object conception of the Laplacian as the "average bending" (HE8). Finally, we proceeded to probe the coordination (see Fig. 10) and understanding of the Laplacian of the temperature and the heat equation (HE10). Throughout this question, we implicitly probed Pre-ii, the schema conception for the Laplacian, gradient, and divergence.

### B. Participants and interviews

We carried out think-aloud interviews with 8 students in the second year of their B.Sc. These students were enrolled in either engineering, physics, or twin (mathematics and physics) majors. The students were recruited on a voluntary basis. They agreed to take part in the interviews after the project was presented to them by the first author. In the first semester of their second year, they had completed a course that covers the 2D heat equation. They had taken exams and received their final course grades in February, and were interviewed in March and April of the same year. All courses are graded out of 20; the lowest grade among the volunteers was 12. Table IV shows each student's major, and the final grade they received on the course. The participants constituted a diverse sample with different performance levels. This allowed us to probe mental constructions of a 'generic' student who passed the course. Not all participants solved all questions. Table V shows a categorization of the students' performance level per question.

Each interview took around 60 minutes. Students were encouraged to verbally express their thoughts while they solved the questions. We prioritized ensuring that students had sufficient time to answer question 3, as it encompasses several key mental constructions, and we aimed to minimize the impact of time constraints on their responses. Therefore, students answered the questions in the order of 1, 3, 2. Each student solved all three questions except for students 1 and 2, who ran out of time and could not solve question 2. Students carried out the interviews using a smart pen which allowed us to record synced handwriting and audio-recordings. Audio-recordings were transcribed clean verbatim to make the text more readable for analysis. Follow-up questions throughout the interviews were used to probe students' responses further.

The first author examined students' responses and looked for evidence of mental constructions question by question and student by student. The interpretations and supporting evidence were discussed among the authors to ensure consistency in the analysis.

TABLE V. List of students participating in interviews and their corresponding course grade.

| Student | Major | Grade /20 |
|---|---|---|
| 1 | Engineering | 17 |
| 2 | Engineering | 17 |
| 3 | Engineering | 15 |
| 4 | Engineering | 14 |
| 5 | Twin | 13 |
| 6 | Physics | 14 |
| 7 | Engineering | 19 |
| 8 | Physics | 12 |

TABLE VI. Participation of students per question.

| Performance | Q1 | Q2 | Q3 |
|---|---|---|---|
| Average lower range <14 | Students 4,5,6,8 | Students 4,5,6,8 | Students 4,5,6,8 |
| Average Upper range <16 | Student 3 | Student 3 | Student 3 |
| Good <20 | Students 1,2,7 | Student 7 | Students 1,2,7 |
| | Total: 8 | 6 | 8 |

## IV. FINDINGS

### A. Temperature gradient and Fourier's law

In this sub-section, we discuss the mental constructions probed by Question 1 (see Fig. 5). We asked students to explain how they can obtain the heat flux density at 2 points on two 2D plates for temperature distributions given on a Cartesian graph. In part a, we explicitly probed object and process conceptions of the temperature gradient (HE4a), and the proportional relationship between the temperature gradient and heat flux density (Pre-i.a). In part b, we explicitly probed process understanding of thermal equilibrium (Pre-i.c). In part c, we probed the object conception of the temperature gradient as a symbolic representation (HE4b). In all parts, we implictly probed the process conception of the temperature gradient as interiorizing the calculation of a spatial rate of change and obtaining first partial derivatives with respect to spatial variables (HE4c) and the object conception of the temperature gradient as the highest spatial rate of change at a point (HE4.d).

We discuss the mental constructions that students showed evidence of in their responses. We start by showing students' understanding of the vectorial, linguistic, and symbolic representations of the temperature gradient (HE4.a). We then proceed to illustrate students' conceptual confusion when interpreting Fourier's law (Pre-i.a) and a time-dependent temperature distribution. Since all students demonstrated a clear understanding of how thermal equilibrium leads to zero heat transfer (Pre-i.c), we do not discuss this aspect in detail.

*1. The temperature gradient*

Five students (students 1, 2, 5, 7, and 8) showed an object understanding of the vector representation of the temperature gradient (HE4.a). Student 1 made a mistake: they mentioned descent instead of ascent. Student 5 immediately mentioned that it is a vector that points in some direction, reasoning about its direction using changes in the temperature surface—consistent with the idea that it points in the direction of steepest ascent. Three students (students 2, 7, and 8) showed evidence of using a process understanding of the temperature gradient (HE4.c) by obtaining first PDs with respect to the spatial variables $x$ and $y$ to obtain a vector representation for the temperature gradient as we see in the following two examples.

Student 7 stated:

> "…because I know like for example, the gradient of a function is something like this $\frac{\partial T(x,y)}{\partial x}$, $\frac{\partial T(x,y)}{\partial y}$,"

and proceeded:

> "... Law of heat conduction is about the temperature next to point P, so that's why you have to do the derivatives to know what the other point (next to P) are expressing itself that influences the heat conduction. So, you say what is the difference in temperature if you go a bit further along the x axis. So, this one means how much does it change on the x axis, how much does it change like this in this direction. For the y axis, it's the same. If you go a bit further to the y axis, how much is it changed there. If I look, it doesn't change on the y axis so it would point downwards ..."

Student 8:

> "...this is a vector quantity that's the y-component is just constant. So, its partial derivative with respect to y will be zero. If it's just along the x and along y is just constant and then the second component is just zero..."

The remaining three students (students 3, 4, and 6) did not show evidence of having an object conception of the vector representation of the temperature gradient. Students 3 and 4 were explicitly reminded by the interviewer that it is a vector pointing in the direction of the steepest ascent.

Six students showed a correct object conception of the symbolic representation of the temperature gradient (HE4.b), with $x$ and $y$ as spatial independent variables. Only students 2 and 3 chose the symbolic representation of the temperature gradient that includes the temporal rate of change of temperature. None of the students showed evidence of an object conception of the linguistic description of the temperature gradient (HE4.d), i.e. they did not spontaneously describe the magnitude of the gradient as the highest spatial rate of change at a point.

*2. Instantaneous nature of Fourier's law*

As stated above, students 2 and 3 chose an incorrect symbolic representation of the temperature gradient that included a temporal rate of change. Student 2 stated:

> "In these two figures, time did not have any effect on the heat. There was a given time and you would not check like previous moments or later, but now, it is not specified that it is at a specific time t. I would say a change of time would affect the temperature so it would affect the heat flux."

They later wrote:
> "In this question, the heat, maybe, depends on the time, so we must also take the partial derivative with respect to time into account. So, I would choose ii."

As stated in Section IV.A.1, this student showed an understanding of the gradient as a calculation of a spatial rate of change of the temperature in a situation where time is fixed (HE4.c). They referred to the temperature distribution changing with time at every instant, which means that both the temperature gradient and the heat flux density are time dependent. This led them to an incorrect choice for a symbolic representation of the temperature gradient. The student's reference to "heat" varying with time appears to reflect a confusion between heat and temperature. We address this conceptual difficulty in more detail in Section IV.B. Student 3 used a similar approach when they stated:

> "… because the temperature changes in time as the (heat) flux flows so the gradient will be different at each point in time."

These two students do not reason with the temperature gradient at a specific instant. Instead, they highlight that the temperature gradient will change with time. They have developed an incorrect object conception of the symbolic representation of the temperature gradient, despite understanding the physical process quite well.

### B. Heat conduction

This section mainly tackles mental constructions probed by Question 2 (Pre-i.b, HE2, and HE3). This question (see Fig. 6) was administered to 6 students out of 8. Students 1 and 2 ran out of time while answering the other questions.

To probe the mental construction related to heat and temperature (Pre-i.b), students were presented with a 2D plate. One boundary, A, was set to be insulated, while another boundary, B, was described as immersed in ice. We asked students to choose among different options the one(s) representing the insulated boundary A. The correct answers for boundary A indicate a correct object understanding of heat (answer i) and a correct process understanding of the quantification of heat related to thermal energy (answer iii). The wrong answers (ii and iv) refer to process and object conceptions of temperature that do not apply to an insulated boundary. Option ii indicates that the quantification of temperature as thermal energy will be constant in time which is not true for an insulated boundary. Similarly, we asked students to choose among four options the one(s) that describe that the boundary B is immersed in ice. The correct answers for boundary B refer to a process understanding of temperature as quantification of thermal energy (answer ii) and an object conception of temperature (answer iv). Similarly, the wrong answers (i and iii) respectively refer to object and process conceptions of heat that do not apply to a boundary immersed in ice.

To probe the object conception of the spatial and time rate of change of the temperature (HE2 and HE3), students were first asked to choose a correct mathematical expression corresponding to "boundary *A* of the plate is insulated" without the graphical representation of the boundary. Later, they were presented with the graphical representation (see Fig. 7), and asked whether they would keep or change their answers.

We start by giving an overview of students' understanding of the concepts of heat and temperature in relation to thermal energy. Then we highlight the challenges students faced when interpreting the insulated boundary condition.

#### 1. Distinguishing heat and temperature

All 6 students correctly chose the options corresponding to an object conception of heat and temperature. This was not the case for the process conception of heat and temperature (Pre-i.b). Differentiating between heat and temperature in their relationship to thermal energy was a challenge. Three students (5, 6, and 7) chose the option reflecting a process understanding of heat while four students (3, 4, 5, and 6) chose the option reflecting a process understanding of temperature. Indeed, Students 5 and 6 demonstrated evidence of both conceptions.

A lack of process conception in relation to thermal energy could be the reason behind the conflation between heat and temperature as shown previously in the answer of student 2 in IV.A.2. The student interprets heat as a quantity that varies with time similar to temperature. However, from a thermodynamics perspective heat is a quantity representing thermal energy transfer.

#### 2. Insulation and rate of change of temperature

In the PGD, we suggest that students need to understand heat flow as a $1^{st}$ partial derivative with respect to a spatial variable (HE2). We hypothesize that students need three key understandings: the relation between heat and thermal energy, Fourier's law, and $1^{st}$ PD.

In Question 2a, students showed no problem in defining insulation as zero heat flow through boundary A. In Question 2c, however, when they were asked to choose a mathematical expression representing that boundary A is insulated, without seeing the graph, four students out of 6 (students 3, 4, 6, and 8) chose a symbolic expression for the temperature gradient with a spatial variable, although not always the correct one (Table VI). Students 5 and 7 chose a symbolic expression for the temperature gradient with a time variable.

These two students, along with student 6 who chose a wrong expression with a spatial variable, corrected their answers when presented with the graphical representation of the temperature distribution at insulated boundary A (Fig. 7).

Table VII. Overview of student answers for parts of question 2 before (column 2) and after (column 3) they were presented the graphs. Verified processes are shown next to each student's answer (columns 4-6).

| Student # | Before graph | After graph | Fourier's law | Heat | 1st PD |
|---|---|---|---|---|---|
| 3 | $\frac{\partial T}{\partial y}$ | $\frac{\partial T}{\partial y}$ | Y | N | Y |
| 4 | $\frac{\partial T}{\partial y}$ | $\frac{\partial T}{\partial y}$ | Y | Y | Y |
| 5 | $\frac{\partial T}{\partial t}$ | $\frac{\partial T}{\partial y}$ | Y | Y | Y |
| 6 | $\frac{\partial T}{\partial x}$ and $\frac{\partial T}{\partial y}$ | $\frac{\partial T}{\partial y}$ | Y | Y | Y |
| 7 | $\frac{\partial T}{\partial t}$ | $\frac{\partial T}{\partial y}$ | Y | Y | Y |
| 8 | $\frac{\partial T}{\partial y}$ | $\frac{\partial T}{\partial y}$ | Y | N | Y |

Student 5 for example argued:

> Interviewer: *"Can you explain physically why (ii) is your definite answer without referring to the graph?"*
>
> Student 5: *"So the temperature distribution needs to make sure that at y = 0, no heat can flow in or out. So, the point infinitesimally next to the point at y = 0 needs to be the same as y = 0 because otherwise they would exchange heat, and because it needs to be insulated, the point right next to it needs to be the same temperature as y = 0. So, the derivative if we go one step to the y will need to be the same because otherwise, they would exchange heat. So, I think that's the way I can explain the best right now."*

Moreover, students showed gaps in their understanding of insulation where they interpreted insulation as maintaining a constant temperature at a boundary. This applies to students with and without an initial correct answer (i.e., before they were shown the graphs). We show examples from students 4, 8 (correct before seeing the graphs) and 6 (incorrect before seeing the graphs) who all showed that they understand insulation as zero heat flow, and all eventually chose a correct symbolic interpretation. Nevertheless, we see evidence they still interpreted insulation at a boundary as constant temperature at that boundary. When answering which expression best represents an insulated boundary in Question 2a, student 4 chose the first three options (see Fig. 6). While choices i and iii are correct and reflect an object and process conception of heat respectively, option ii would logically imply that this student considers the temperature at boundary A to be constant when it is insulated. Indeed, they elaborated and wrote "$T(x, 0) = constant$" and "$\frac{dT}{dy} = 0$" in answer to part c when asked to choose a symbolic expression representing the insulation.

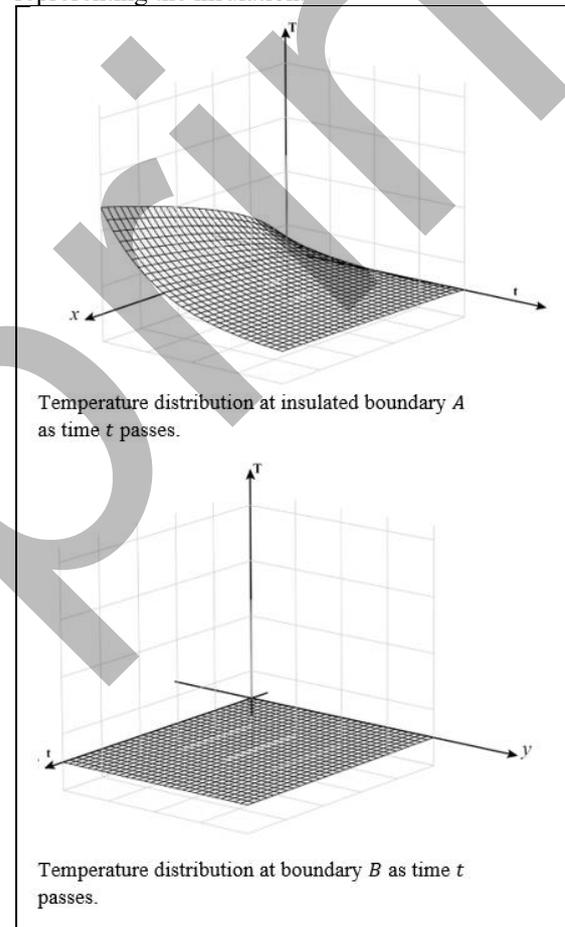

FIG. 7. Temperature distribution at boundaries in Question 2.

Student 6 was unsure if the insulation also means no heat flows parallel to the boundary:

> *"The physical description of A is that there's no heat flow, but don't know whether that heat flow can also just be like parallel to the boundary."*

They originally decided that both $\frac{\partial T}{\partial x} = 0$ and $\frac{\partial T}{\partial y} = 0$. After they were given the graphs, they correctly argued

> *"I think I'll have to change my $\frac{\partial T}{\partial x}$ because of course at any point in time the temperature does not have to be constant across the board [i.e., the boundary]."*

They changed their answer into only $\frac{\partial T}{\partial y} = 0$, which is correct. However, student 6 later stated:

> *"The temperature cannot vary according to y at the boundary A."*

We see student 6 implicitly interpreting $\frac{\partial T}{\partial y} = 0$ as constant temperature instead of zero heat flow or zero temperature gradient in the $y$ direction. The temperature along $y$ at boundary A can still vary with respect to the spatial variable $x$ and with respect to time $t$.

Student 6 seems to continue thinking that a partial derivative being equal to zero is equivalent to the function being constant, whereas in the case of $\frac{\partial T}{\partial y} = 0$ it only means that the tangent lines in the $y$-direction are horizontal for points along boundary A. The temperature along $y$ at boundary A can still vary with respect to the spatial variable $x$ (and with respect to time $t$). This inhibits a physical interpretation in terms of zero heat flow or zero temperature gradient.

Student 8 (having been shown the graphs):

> *"So I chose for part c option two, which says that the partial derivative of the temperature with respect to y is 0. If I've chosen the other options, so at boundary [A] the derivative with respect to the x, there should be a flat surface. That's not the case, and I guess it's also for the time component. If I look at it correctly, it doesn't seem to be constant 0. So, I guess still stick to my option 2 in question."*

Student 8 stated there should be a flat surface (constant temperature) in the given graphical representation if it was insulated at the $x$ axis boundary.

It seems students use inconsistent terminology that reflects confusion in their understanding of insulation (students 4 and 8) or insufficient phrasing of reasoning (student 6). All students who answered the question eventually interpreted the insulated boundary correctly as $\frac{\partial T}{\partial y} = 0$. However, reasoning with the predicted three processes does not seem sufficient to make the students' explanations clear and sufficient to fully ensure that students distinguish between the boundary condition representing insulation—where the temperature gradient normal to the boundary is zero—and the incorrect interpretation of insulation as implying a constant temperature along the boundary. We hypothesize that this potential misinterpretation could stem from a failure of interplay of mathematics and physics. Even though students can interpret insulation correctly as zero heat flow, their interpretation of $\frac{\partial T}{\partial y} = 0$ as constant temperature may be based on a correct mathematical statement (if the derivative of a function of one variable is 0 along an interval, then this function is constant on that interval) which is generalized incorrectly (if the partial derivative of a function with respect to one of its variables is zero, then this function is also constant with respect to the other independent variables).

### C. Laplacian of the temperature and the 2D heat equation

This section mainly tackles the mental constructions probed by question 3 (Pre-iv and HE5-HE10). In this sub-section, we describe students' process conceptions of the Laplacian of the temperature as the divergence of the temperature gradient (HE5) in IV.C.1 and as the sum of 2nd partial derivatives (HE7) as well as their object conception of the Laplacian as "average bending" along $x$ and $y$ (HE8) in IV.C.2. We show students' object conceptions of spatial concavity along one spatial variable (HE6). We illustrate how coordination of HE5 and HE7 (HE9) could be a mechanism that enhances understanding of how mathematical reasoning relates to heat flow in IV.C.3 and could be a (challenging) pathway for the interpretation of the Laplacian of the temperature and therefore an interpretation of the 2D heat equation (HE10) in IV.C.4.

Table VII gives an overview of students' understanding regarding these processes, their coordination and their encapsulation to interpret the Laplacian of the temperature.

#### 1. The Laplacian as the divergence of the temperature gradient

To probe the students' process conceptions of the Laplacian of the temperature as the divergence of the temperature gradient (HE5), we presented them with 2 gradient vector fields: (a) gradient vectors with only one non-zero component (along the $x$-axis) and (b) gradient vectors with two non-zero components (see Fig. 8). They were asked to identify the Laplacian at points $P$, $Q$, $M$, and $N$ as positive, negative, or zero.

As shown in Table VII, extending students' understanding from gradient vector fields with one non-zero component (points $P$ and $Q$) to more complex gradient vector fields with two non-zero components (points $M$ and $N$) is particularly challenging. We elaborate on this finding.

For gradient vector fields with one non-zero component, five students (students 2, 4, 5, 6, 8) out of eight correctly obtained the sign of the Laplacian of the temperature at points $P$ and $Q$. We also included student 1 because he was able to reason with the Laplacian as the divergence of the gradient although he did not arrive at the correct sign due to inconsistency in the use of terminology. First, we show quotes from students with correct signs then we proceed to discuss student 1 in detail.

Table VIII. Overview of students' understanding of the Laplacian of the temperature.

| Student | HE5 T Laplacian as $\vec{\nabla} \cdot \vec{\nabla}T$) | | HE7 T Laplacian as sum of spatial 2nd PDs | HE9 Coordination | HE9 $\nabla^2 T$ interpretation |
|---|---|---|---|---|---|
| | $\vec{\nabla}T$ 1 component | $\vec{\nabla}T$ 2 components | | | |
| 1 | ✓ | ✗ | ✓ | ✗ | ✗ |
| 2 | ✓ | ✓ | ✓ | ✗ | ✓ |
| 3 | ✗ | ✗ | ✗ | ✗ | ✗ |
| 4 | ✓ | ✗ | ✓ | ✗ | ✓ |
| 5 | ✓ | ✓ | ✓ | ✓ | ✓ |
| 6 | ✓ | ✗ | ✓ | ✓ | ✓ |
| 7 | ✗ | ✗ | ✗ | ✗ | ✗ |
| 8 | ✓ | ✗ | ✓ | ✗ | ✗ |

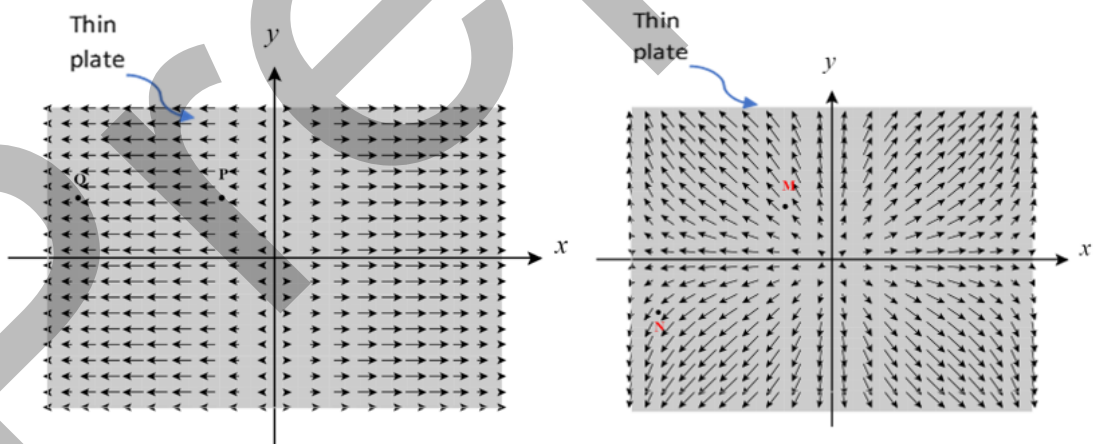

The heat equation describes heat flow in a thin plate. It can be expressed in two equivalent ways:

$$\frac{\partial T}{\partial t} = \alpha \left(\frac{\partial^2 T}{\partial x^2} + \frac{\partial^2 T}{\partial y^2}\right) \text{ or } \frac{\partial T}{\partial t} = \alpha \vec{\nabla} \cdot \vec{\nabla} T$$

$T$ is the temperature at a point $(x, y)$ at time $t$. $\alpha$ is the thermal diffusivity of the thin plate. $\frac{\partial T}{\partial t} = \alpha \Delta T$ is another compact way to write the heat equation. Given below two gradient fields describing temperature distribution on two thin rectangular plate $A$ & $B$ in figures 1 and 2 below at instant $t = t_1$.

Fig 1: Gradient vector field of temperature distribution on plate $A$ at $t = t_1$.

Fig 2: Gradient vector field of temperature distribution on plate $B$ at $t = t_1$.

a. Identify the temperature Laplacian $\Delta T$ at points $P, Q, M, N$ as positive, negative, or zero. Explain your reasoning.

FIG. 8. Question 3-part a, showing temperature gradient vector fields with marked points for analyzing $\Delta T$.

Student 2: *"... you can see that the arrows are going from very small to quite big and then back to small. Hmm, like in point P, this will lead to a positive Laplacian and in Q, a decrease of magnitude. So, you would get a negative Laplacian."*

Student 6: *"There are more arrows pointing outwards of P than inwards. You can kind of see that the length of the arrow changes according to the x axis. Q should be negative, and positive at P, like, the direction of the arrows, is also kind of important."*

By contrast, only two students (students 2 and 5) out of eight correctly obtained the sign of the Laplacian of the temperature at points $M$ and $N$ for the gradient vector field with two non-zero components. Although students 4, 6, and 8 reasoned correctly with the divergence of the temperature gradient in the first situation, they did not extend this understanding to the 2D case. For example, student 6, who used both integral and differential approaches in the first case, wrote:

Student 6: *"… then since for M and N just always outflowing should just be positive"*

Student 1 identified the Laplacian of the temperature as the divergence of the temperature gradient but inconsistently applied this concept, leading to an incorrect sign for the Laplacian of the temperature for the points on the plate with temperature gradient vectors with one non-zero component.

Student 1: *" more vectors are travelling away from it (point P) than entering it, so I believe the temperature is going down so delta T (Laplacian of T) is negative, and for Q we've got more arrows arriving than going away so that (Laplacian of T) must be greater than zero, because the temperature gradient goes from high to low…"*

They recognize that the Laplacian of the temperature can be considered as the divergence of the temperature gradient by applying the integral approach for gradient vectors "travelling away from" and "entering" (a small box around) point $P$. However, this student is inconsistent in the conceptual interpretation of the Laplacian of the temperature. They might be making three distinct mistakes. The first one is misinterpreting an incoming temperature gradient arrow as indicating incoming heat and an outward arrow as outgoing heat. This leads to obtaining a wrong sign for change in temperature which the student then mistakenly identifies as the Laplacian of the temperature — the second mistake. Then they proceed to indicate how the gradient vector field itself is changing in the vicinity of point $Q$ "*because the temperature gradient goes from high to low*", but they do not take into consideration that the temperature gradient vectors point towards the negative $x$-axis. They are following the direction of the arrows instead of the $x$-axis—a third mistake.

However, this student did not follow the same approach in the case with two non-zero components:

Student 1: *"Now we have N, but the vectors are a bit more curl, I would say more in than out so it is +. For M I think it is zero, I think, I have no idea …"*

They continued even when they were prompted by the interviewer to follow the same approach:

Interviewer: *"... so for M you used a different reasoning or am I wrong?"*

Student 1: *"...if you were to zoom in closer and closer ... at M ... it [the vectors] would be the same. ... I think if you were to zoom in, it [the vectors] would be the same going in as going out."*

Interviewer: *"So, you think if you were to zoom in on M, the vectors coming in will be equal to the vectors coming out and it is not the same for N?"*

Student 1: *"The vectors seem to be getting smaller to the middle [seems as they are referring to the whole plate] so I think there will be a difference. I think more is leaving than going in [referring to M] they seem small and then get bigger and bigger over time. So, I think for M, more going out so change for temperature would be smaller (than zero). I think same is for N then."*

[Student proceeds to write: "N: <0. M: <0"]

Interviewer: *"Both have the same sign, or did I read this wrong?"*
Student 1: *"Yes, yes, both of them have the same sign."*

Students 1 and 6 are examples of how extending an understanding of the Laplacian of the temperature as the divergence of the temperature gradient is a challenge. Student 6 reasons with the divergence of the whole gradient field instead of at a point. Student 1 seems to do something similar when questioned by the interviewer.

2. *The Laplacian as the sum of spatial 2$^{nd}$ PDs and average spatial bending*

To probe the process conception of the Laplacian of the temperature as the sum of 2$^{nd}$ PDs (HE7) and the object conception as "average bending" along $x$ and $y$ (HE8), students were given a graph of a temperature distribution on a 2D plate. In part e of question 3, students were asked to indicate at which point(s) the Laplacian of the temperature is positive. Students 1, 2, 5, 6, and 8 reasoned correctly and also showed evidence that they have encapsulated this process into an object understanding of the Laplacian of the temperature as "average bending" along $x$ and $y$. We showcase one example from student 8 (and will return to students 1 and 2 later).

Student 8: *"Laplacian is positive when these second order derivative with x and y are both positive …"*
Interviewer: *"That applies to which point?"*
Student 8: *"J."*
Interviewer: *"What do you think would be the Laplacian at K…"*
Student 8: *"... if the curvature is somewhat ... if the shape is somewhat symmetric. I guess the 2$^{nd}$ partial derivatives with respect to $x$ and $y$ will be equal but opposite I guess and then they add up to zero."*

This student refers to "symmetry" of curvature when reasoning with the Laplacian of the temperature at point $K$, a sign that they have encapsulated the sum of spatial 2$^{nd}$ PDs and is now reasoning with "average bending".

However, students 3 and 7 did not show evidence of these mental constructions. Student 4 could only identify the equivalency between the Laplacian of the temperature and the sum of spatial 2$^{nd}$ PDs, however they could not obtain the correct sign. From their answers to parts b and c of question 3, in which we probed parts of the schema of 2$^{nd}$ PDs (Pre-iv), it is clear that these students lack this important prerequisite. Students were asked how they would represent the second partial derivative of the temperature with respect to $x$, graphically (part b) and linguistically (part c).

We show a quote from student 3.

Student 3: *"Just a graph"* [pauses] *"I really don't know"*
Interviewer: *"Okay, let's say you have a curve, how would you represent the 2$^{nd}$ partial derivative"*
Student 3: 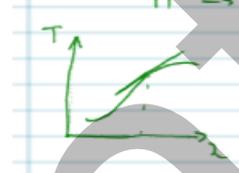

*"… At a given point, this is the 1$^{st}$"* [proceeds to draw tangent at point] *"and then the 2$^{nd}$"* [pauses] *"what was it?"*
Interviewer: *"So, if the 1$^{st}$ partial derivative at this point is the slope of the tangent, what would be the 2$^{nd}$ partial derivative?"*
Student 3: *"It's just the slope, but not of this graph, if you take it (slope) at every point, you get a new graph, and it is the slope of the 2$^{nd}$ graph."*

This quote shows that the student has a correct formal conception of the second derivative as the derivative of the first derivative. However, in part d, when asked to obtain the sign of the 2$^{nd}$ PD at points $J$ and $K$, student 3 stated that both 2$^{nd}$ PDs for $J$ and $K$ were zero. The student seems to obtain the derivative of the value of the 1$^{st}$ PD (zero) instead of comparing two slopes of tangents in the vicinity of a given point. Student 4 similarly clearly states:

Student 4: *"... so the 1$^{st}$ partial derivative would give a tangent plane which in J would be a constant plane, so $\frac{\partial^2 T}{\partial x^2} > 0$ is certainly not J."*

We see how an incomplete schema understanding of the 2$^{nd}$ PD could affect students' understanding of the Laplacian of the temperature and could prevent them from obtaining the correct sign at a point.

b. How do you represent $\frac{\partial^2 T}{\partial x^2}$ graphically?
c. Which of the following best interprets $\frac{\partial^2 T}{\partial x^2}$ physically in regard to temperature distribution on the thin plate?
   i. Time rate of change of temperature.
   ii. Spatial rate of change of temperature.
   iii. How time rate of change of temperature varies with respect to time.
   iv. How spatial rate of change of temperature varies with respect to a spatial variable.

A thin rectangular plate $C$ with temperature distribution shown in figure 3 from different perspectives at $t = t_1$. $J$ and $K$ are points on the plate.

Perspective 1

Perspective 2

Fig 3: Temperature distribution on thin plate $C$ at $t = t_1$.

d. Which point(s) ($J$, $K$, both, or none) depict(s):
   i. $\frac{\partial^2 T}{\partial x^2} > 0$
   ii. $\frac{\partial^2 T}{\partial y^2} < 0$?
   iii. $\frac{\partial^2 T}{\partial x^2} < 0$
   iv. $\frac{\partial^2 T}{\partial y^2} > 0$
e. Which point(s) depicts a temperature Laplacian $\Delta T > 0$?

FIG. 9. Question 3 parts b-e, showing how we probed mental constructions Pre-iv, HE6, HE7 and HE8.

Students 1 and 2 show a lack of consistency when engaging with the second partial derivative schema. When answering part b, they seemed not to activate their schema of $2^{nd}$ PD (e.g., by explicitly referring to *concavity* at a point).

Student 1: *"no idea ... I guess the. Uhm .... I guess like the first (partial derivative) would be the change of temperature when you travel only in the x direction, and this ($2^{nd}$ PD) would be the acceleration ... No idea how I would do that graphically though."*

Student 2: *"You can always draw the first derivative and then the second."*
*"I would try to draw the first derivative by like looking if it's increasing or decreasing ..."*
*"But it would not be like very accurate, just the positive and negative parts (of the function) would be visual because those points are always easy to check and then I would again take the derivative of the derivative I drew."*

However, students 1 and 2 showed evidence that they have developed a process and object understanding of the Laplacian of the temperature as the sum of spatial $2^{nd}$ PDs and "average bending" in their answers in parts d and e. They were able to reason with $2^{nd}$ PD at point $J$ and $K$ by referring to how a slope is changing in the vicinity of a point which we considered a process understanding of $2^{nd}$ PD.

Student 1: *"... so we have a minimum and that's in the y direction so that only applies for J. Well, I always remember that like in high school I was told if the first derivative of some function with respect to x is equal to 0 then you can go look at the second derivative, and if that's negative you're at the maximum because like you're going from like you're decelerating like you're going down the hill and if it's positive, you're at a minimum. So, I think this also applies for higher dimensions."*

Student 2: *"It seems to me that in point J, I would be in a minimum, so the first derivative would be 0, but not the second one. The slope of the first derivative would be 0."*
*"For point K, this the (first partial) derivative (along x) will be a constant function. This (referring to concavity along x) is increasing, so the second partial derivative of temperature*

*through x in point K would be greater than 0. Same for J."*

Inconsistent activation of the second partial derivative schema for some students, and lack of development of this schema for other students suggest that students need support in their understanding. We elaborate more on this in IV.D.

### 3. Coordination: Possible Mechanism to improve understanding

In this subsection, we provide evidence of how coordination of the Laplacian processes HE5 and HE7 could enhance understanding of heat flow as a quantity involving a first partial derivative with respect to a spatial variable.

In parts f-i of question 3 (see Fig. 10), we probed the coordination of two processes: the Laplacian of the temperature as the divergence of the temperature gradient and as the sum of $2^{nd}$ spatial PDs (HE9). We also probed mental constructions related to the interpretation of the Laplacian of the temperature (HE9) and the 2D heat equation (HE10).

As shown in Table VII, students 5 and 6 carried out the coordination successfully. When asked to interpret the Laplacian of the temperature (part h), student 5 answered as follows:

Student 5: *"If you look at a point and we say we take the temperature Laplacian at that point, we see what is the average temperature in the region and how it is changing, what part is hotter, what part is colder. So, we take the small area in the two dimensions around the point that you're trying to look at and we will see, yeah, what part is hotter and to where is it hotter or not... [The Laplacian of the temperature] is the change [difference] in temperature at a point and around it at every point."*

Student 5 starts by showing a process-like conception of the Laplacian of the temperature when comparing the temperature at a point to the temperature of points surrounding it ("what part is hotter, what part is colder"). Towards the end, they show more of an object-like conception when referring to the temperature difference between a point and its surroundings as an entity. This is later clarified when the student refers to concavity at a point when interpreting the 2D heat equation.

In their answer to part i, they continue to draw a 1D cross-sectional representation of the temperature along the $x$-axis as follows and state:

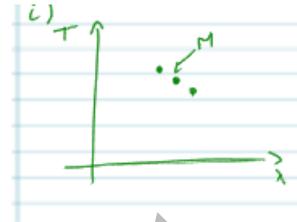

*"The hotter one will flow to M, but M will flow equally as much to the lower one, so M will be the same ... It's a nice line okay. So, the second derivative is zero, it isn't convex or concave."*

They later draw a concave and a convex curve, draws a slope at neighboring points, and refers to heat flow from/to the middle point.

*"This one (upper point on concave arc) will be higher than this one is (lower point of concave arc), so this one (top) will give more energy than this one (bottom) will take. So, this one (top) will give more than this one (middle) needs to give, so it (middle) will get more than it needs to give, so it(middle) will rise."*

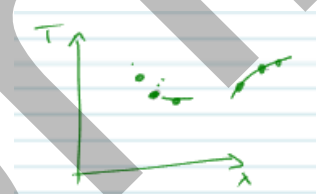

Regarding coordination in part f, student 5 incorrectly chose B as a gradient vector field that matches the temperature distribution around J.

Student 5: *"So the Laplacian would be positive. How would heat flow? It would flow in all directions into J. Yes. So B ..."*
Interviewer: *"Ok, and if you were to use your answer for part d, would you be able to give a supporting argument for your choice of B?"*
Student 5: *"... So both of them (2nd PDs) are positive. Laplacian is always positive and positives mean that it would be a sink, and this feels like a sink of the vector fields."*
Interviewer: *"... in your answer to part e you indicated the Laplacian at point J is positive, can you reason with your answer to part d where you take each second partial derivative separately?"*

f. Using your answer for part d, which of the following depicts the Laplacian at $J$. Explain your reasoning.

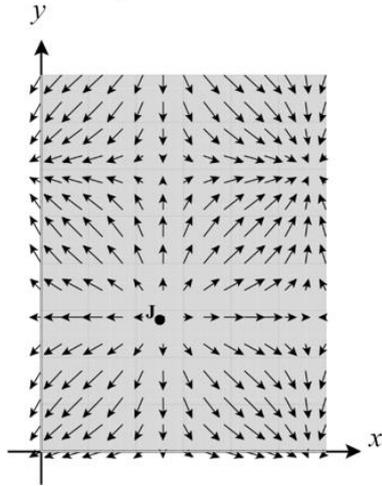

A

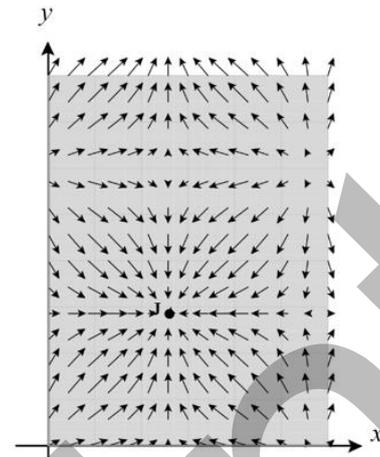

B

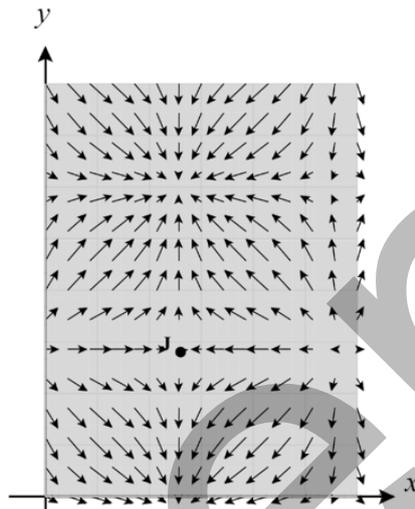

C

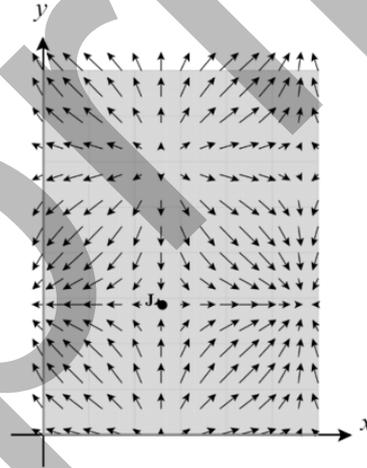

D

g. Temperature Laplacian $\Delta T$ at M in figure 2 is equivalent to: (Choose all that apply)

| i. | $\frac{\partial^2 T}{\partial x^2} > 0$ |
| --- | --- |
| ii. | $\frac{\partial^2 T}{\partial x^2} + \frac{\partial^2 T}{\partial y^2} < 0$ |
| iii. | $\frac{\partial^2 T}{\partial x^2} < 0$ |
| iv. | $\frac{\partial^2 T}{\partial x^2} + \frac{\partial^2 T}{\partial y^2} > 0$ |
| v. | $\frac{\partial^2 T}{\partial y^2} < 0$ |
| vi. | $\frac{\partial^2 T}{\partial y^2} > 0$ |

h. How can you interpret $\Delta T$ physically?
i. How can you interpret $\frac{\partial T}{\partial t} = \alpha \, \Delta T$ physically?

FIG. 10. Question 3 parts f-i, showing how we probed students' coordination of the Laplacian of the temperature as the divergence of the gradient of the temperature and as the sum of two spatial second partial derivatives.

Student 5: *"At point J, I would draw an imaginary line parallel to the y axis and I'm going to see how the arrows flowing into it would change and we can see that it goes from negative (arrow above J) to a positive (arrow below J), so it means it has gone up, so it needs to be bigger than zero. We draw a line parallel to x going through J and we can see that it would go again from positive (left of J) to negative (right of J). No, from negative. Yeah. Now I see it. I have flipped them around ..."*

We see that coordination of HE5 and HE7 allowed student 5 to realize that they made an error and improve their reasoning to obtain the divergence of the temperature gradient. This student reasoned physically by stating that heat should flow into J since the Laplacian of the temperature is positive. However, they seem to be interpreting the gradient vector field as heat flux density vectors, and the interviewer corrects this for them.

Interviewer: *"All of the four choices are gradient vector fields"*

Student 5: *"OK, OK, OK. And the gradient is the opposite of the heat flow as we saw in the previous questions, because $\vec{q}$ is minus the gradient. Yeah, so then indeed it would be A."*

In part g, the same student attempts to coordinate again between the Laplacian at point M as the divergence of the temperature gradient and as the sum of spatial 2$^{nd}$ PDS.

Student 5: *"I answered at M that the Laplacian was positive, so the sum of these two (2nd PDs) would also be positive, but then the individual aspects, we need to take a closer look. So, we just take the derivative of this one (gradient vector M) and that is obviously changing in both x and y, OK. So, I think both x and y are positive. So, i, iv and vi."*

After this reasoning, the student realizes an error they made in their response to part a of question 3. They indicated a wrong value for the Laplacian of the temperature and followed an approach where they looked not only how the vectors are changing at a point, but also how their rate of change is also changing, in the sense they were applying a 2$^{nd}$ PD to the gradient vectors.

Previously and before coordination, student 5 incorrectly indicated that the Laplacian of the temperature for both P and Q (fig. 8) would be zero.

*"I'm just going to go with the sum of the two double derivatives because I think that's easier and clear in my head. I'm going to say for Q and P, I think I would say is 0 ..."*

When referring to point M before coordination they state:

*"... and I'm thinking how is that line changing and how fast is the change happening as we go along the diagonal."*

After coordination they state:

*"... and with that **incredible insight**, I think that I can go back to all these (points P, Q, M and N) and handle them a bit smarter. I don't need to see the second derivative, but just the first derivative if that line (vector) is changing or not, and with that P would be positive, Q would be negative ..."*

In summary, we see that coordination allowed student 5 to come to a correct understanding. They were then also able to give a form of interpretation for the Laplacian of the temperature.

4. *Coordination: Possible yet challenging pathway for interpretation of temperature Laplacian*

In this sub-section, we demonstrate students' interpretation of the Laplacian of the temperature in relation to their attempts to coordinate the processes predicted in the PGD. We give evidence of how it is a possible pathway for the interpretation of the Laplacian of the temperature, albeit a challenging one. As shown in Table VII, the two students who were successful in coordinating also gave an interpretation for the Laplacian of the temperature. Student 5 quotes were shown in IV.3. We show quotes from student 6 who answered part h of question 3 as follows:

*"It is basically how much the rate of separate change changes so. If you were to think about it physically it's more like if the, the curvature ohh no, how steeper the temperature difference is at two different points, how, the greater the temperature will change according to time."*

The other students were not successful in performing the predicted coordination. Two of these (students 3 and 7) lacked both processes to perform the coordination. They also were not able to provide an interpretation of the Laplacian of the temperature.

| Interviewer: | "Can you give a physical interpretation of the Laplacian?" |
| --- | --- |
| Student 3: | "mmm... now I cannot say immediately... no I don't." |
| Student 7: | "..No, no, not at all." |

Students 1 and 8 carried out the processes partially but were not successful in coordinating them. These students did not give an interpretation for the Laplacian of the temperature. Supporting students in instruction to carry out these processes completely may help them to carry out the coordination and therefore be able to interpret the Laplacian of instruction.

On the other hand, students 2 and 4 interpreted the Laplacian of the temperature. None of them was successful in carrying out the coordination. Specifically, student 2 showed evidence of predicted processes as shown in table VII, they were not successful in executing the coordination. This suggests that successfully coordinating these processes may be key to understanding the concept. Students 2 and 4 did interpret the Laplacian of the temperature as follows when answering part h.

| Student 2: | "... a flux is a vector, but a Laplacian is a scalar, so I am not sure how to express the conversion from vector to scalar, but I would say it is a measurement for the change in heat flux but with no respect to direction." |
| --- | --- |
| Student 4: | "If you take the Laplacian at the temperature exactly, then I think the Laplacian of the temperature tells how much the flux of the heat changes …" |

### D. Key Insights and Refined Genetic Decomposition

In this sub-section we describe the key insights we can take from students' engagement with the preliminary genetic decomposition. We will comment on our PGD in two ways. We describe and argue validated and reviewed mental constructions. If students show evidence of forming a mental construction, it is validated. If students show evidence of difficulty or make different mental constructions, then the predicted mental construction must be reviewed leading to a refined genetic decomposition.

As for the prerequisites of the PGD, we consider all parts as validated since students show evidence of their reliance on these foundational concepts as described in the preliminary genetic decomposition. These include schema conceptions of heat conduction (e.g., process understandings of Fourier's law, temperature, heat, and thermal equilibrium), vector calculus concepts (e.g., gradient, divergence, Laplacian), partial derivatives, multivariable functions, and spatial reasoning in three dimensions. While these constructions appear to be internalized overall, students still require support with certain aspects. The latter are not considered needed refinements to the PGD per se, but we interpret them rather as indications to areas where instructional support could be embedded in future learning materials.

In particular, students need support to deeply understand second partial derivatives (Pre-iv). We will address this in subsequent sections.

As for the PGD of the heat equation (HE), we give an overview in Table VIII about mental constructions that were validated or shown to need refinement. We consider HE3 validated because students treat $\frac{\partial T}{\partial t}$ as an object. They flexibly apply the first partial derivative schema to the variables of temperature and time. For example, student 7 states when answering part c of question 3:

"This one $[\frac{\partial T}{\partial t} = 0]$ means that the temperature doesn't change in time."

We consider HE5 validated because students show evidence of a process conception of the Laplacian as the divergence of the temperature gradient as previously discussed in Table VII. These students go further in their interpretation of the Laplacian as indicated earlier in Table VII compared to those who cannot carry out this mental construction. We consider HE6 validated since students 1, 2, 5, 6 and 8 chose the correct option in part c where this mental construction is probed. The same students were able at a later stage in part e of question 3 to reason with the Laplacian of the temperature as "average bending". We consider HE7 and HE8 validated since students are able to carry out these predicted mental constructions in part e of question 3. Students 1, 2, 5, 6 and 8 gave the correct sign for the Laplacian in part e. As stated in IV.C.2, they referred to "average bending" when discussing the Laplacian in their answer. For example, student 8 states:

"*If the curvature is somewhat* [pauses] *If the shape is somewhat symmetric, I guess the second partial derivatives with respect to x and y will be equal but opposite, I guess, and then they add up to zero.*"

We consider HE9 validated since, as previously discussed in IV.C.3 and IV.C.4, two students show evidence of carrying out this mental construction. These students were able to interpret the Laplacian of the temperature. We consider HE10 also validated. We give two students' quotes as evidence.

Student 2: *"The change of heat flux would in a way relate to the time dependency of the temperature."*

Student 6: *"The steeper the temperature difference is at two different points, the greater the temperature will change according to time."*

In what follows, we describe and argue the refinements of the PGD that we propose to produce a reviewed genetic decomposition, and highlight mental constructions where students need support which could be embedded in learning material at a later stage.

Table IX. Overview of reviewed and validated mental construction of concepts directly tied to the structure of the two-dimensional heat equation.

| HE mental construction | Reviewed/validated |
| --- | --- |
| HE1. A schema understanding of temperature distribution which includes | Reviewed |
| HE2. An object conception of heat flow as a 1$^{st}$ PD with respect to a spatial variable | Reviewed |
| HE3. An object conception of temperature time rate of variation as $dT/dt$ arising from thematizing the schema first partial derivative to include time as an independent variable (pre-requisite iii) | Validated |
| HE4. A schema conception of the temperature gradient | Reviewed |
| HE5. A process conception of temperature Laplacian as divergence of gradient of $T$: student interiorizes how grad T is affected by the divergence operator without performing calculations. | Validated |
| HE6. An object conception of 1D spatial concavity as how spatial rate of change varies with respect to a spatial variable. | Validated |
| HE7. A process conception of temperature Laplacian as sum of 2$^{nd}$ partial derivatives of $T$ with respect to independent spatial variables. | Validated |
| HE8. An object conception of the Laplacian of $T$ as "average bending" along 2 spatial independent variables at a point | Validated |
| HE9. Coordinating process HE5 and process HE7 to produce a new object that encapsulates the physical interpretation of a temperature Laplacian regardless of the representation; how $T$ at a point compares to its surrounding. | Validated |
| HE10. De-encapsulate the symbolic representation of the heat equation object into process conception of the heat equation. | Validated |

### 1. HE1 and HE2: Schema conception of Temperature Distribution and Object conception of Heat Flow

As summarized in Table VIII, the mental processes from our prediction described in HE2 are not enough to have a precise understanding of heat flow as related to (in this case) $\partial T/\partial y$. More specifically, three students showed evidence of having the three predicted processes: Fourier's law, heat flow, and 1st PD (see Table VII), but did not develop an object conception of heat flow as proportional to the spatial rate of change of the temperature with respect to the correct spatial variable. Student 6 initially seemed to lack a physical meaning for insulation in 2D. Students 5 and 7 interpreted zero heat flow as constant temperature with respect to time before they were shown temperature distribution graphs of the boundaries. Therefore, our prediction (HE2) is not enough to tackle this difficulty, and the genetic decomposition needs to be refined to cater for this. We suggest that students need to distinguish situations of "zero heat flow/flux" and "constant temperature". Students need to reflect on how both, none, or either could be true depending on the boundary conditions. A schema understanding of the temperature distribution (HE1) that includes different conditions of heat transfer is therefore necessary. We predict in a refined version of the temperature distribution schema (see Fig. 11) that students need to have as subschemas:
1. Steady-state temperature distribution;
2. Time-dependent temperature distribution;
3. Temperature distribution with thermal equilibrium.

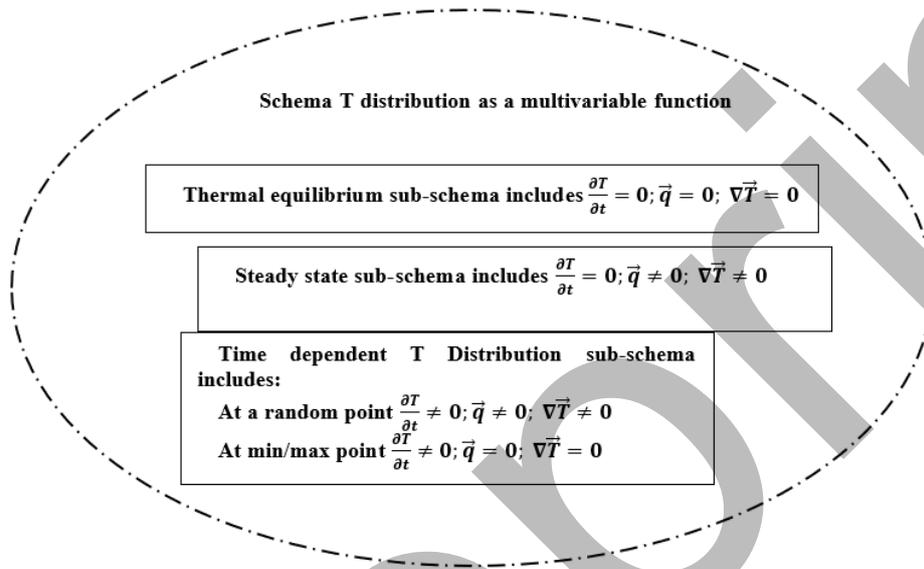

FIG. 11. Refinement of temperature distribution schema HE1.

Each sub-schema includes conditions for an object understanding of the heat flux density $\vec{q}$, the temperature gradient $\vec{\nabla}T$, and the temporal rate of change of the temperature $\frac{\partial T}{\partial t}$.

### 2. HE4: Schema conception of Temperature Gradient

As stated earlier, some students seem to encapsulate a process conception of the temperature gradient as the calculation of a spatial rate into the object conception of a vector representation of the temperature gradient. In addition, we gave evidence of students who reason with the temperature gradient at multiple instants of time, causing them to choose a symbolic representation for the temperature gradient that includes time as an independent variable.

Therefore, supporting students in encapsulating the process conception of the temperature gradient as the calculation of a spatial rate could lead to an object understanding of the temperature gradient as a vector representation. Stressing the instantaneous nature of Fourier's law and how the temperature gradient describes the spatial rate of change of the temperature at a specific instant and how in a time-dependent temperature distribution, we refer to the temperature gradient at a specific instant, might help students to correctly represent it symbolically.

### 3. HE7 and Pre-iv: Process conception of the Laplacian of the Temperature and Schema conception of second partial derivatives

A lack of schema understanding of the second partial derivative (Pre-iv-a,b) can hinder students' ability to correctly interpret the Laplacian of the temperature at a point as the sum of second (spatial) partial derivatives (HE7), and as "average bending" (HE8). Two students showed evidence of developing a process and object understanding of the Laplacian of the temperature as the sum of spatial 2nd PDs and "average bending" consecutively in parts d and e of question 3. However, when answering part b of question 3, they seemed unable to activate their schema of 2nd PD by explicitly referring to *concavity* at a point. The same students were later able to reason with 2nd PDs at points *J* and *K* by referring to how a slope is changing in the vicinity of a point, which we considered a process understanding of 2nd PD.

Therefore, we consider that students require guidance to build a more robust schema conception of the second partial derivative which could lead to a better understanding of the Laplacian of the temperature.

## V. CONCLUSION

In this study, we tested a hypothetical learning trajectory for the 2D heat equation. To do so, we used a main tool from the APOS theoretical framework, the preliminary genetic decomposition. We designed interview questions to probe specific mental constructions within the hypothesized learning trajectory. We carried out think-aloud interviews with 8 students in the second year of their B.Sc. These students were enrolled in either engineering, physics, or twin (mathematics and physics) majors. The participants constituted a diverse sample with different performance levels. This allowed us to probe mental constructions of a 'generic' student who passed the course. Not all participants solved all questions.

Several students demonstrated an object understanding of the temperature gradient as a vector and correctly used symbolic representations involving spatial variables. Some also showed a process understanding by computing partial derivatives to construct the gradient. Some students incorrectly incorporated time into the symbolic representation of the temperature gradient. Although they recognized that temperature varies with time, they did not reason with the gradient at a fixed moment. This indicates a flawed object conception of the gradient of the temperature, despite a reasonable grasp of the underlying physical process. In our previous study [31], students struggled to accurately describe steepness at a point in a Cartesian graph when interpreting the gradient concept. This difficulty, along with students' incorrect interpretations of the temperature gradient, reflects a shared conceptual challenge—namely, the inability to encapsulate processes into a coherent and accurate object.

While students generally identified insulation correctly as zero heat flow, many struggled to translate this into an accurate symbolic representation. Some interpreted insulation as constant temperature at a boundary, revealing confusion between zero temperature gradient and constant temperature. Even when they eventually selected the correct expression ($\frac{\partial T}{\partial y} = 0$), their explanations often reflected imprecise reasoning. Previous literature [26] highlights how graphical reasoning can support the blending of mathematics and physics particularly in relation to linking partial derivatives and their physical meaning. Our findings add further nuance. Equating $\frac{\partial T}{\partial y} = 0$ and constant temperature at the boundary illustrates that the challenge is not only relating heat flow to a spatial derivative. Helping students develop a more refined understanding about partial derivatives within a schema of temperature distribution as shown in Fig. 11 may be crucial.

Students found it significantly more difficult to reason about the Laplacian of the temperature in gradient fields with two non-zero components compared to simpler cases. While several students correctly determined the sign of the Laplacian in one-dimensional gradient fields, few extended this reasoning to more complex two-dimensional cases. This suggests that interpreting the Laplacian as the divergence of the temperature gradient remains a conceptual challenge, particularly when students shift to more spatially complex representations. Klein et al. [32] argue that visual cues can enhance students visual understanding of divergence.

Some students successfully reasoned with the Laplacian of the temperature as both the sum of second partial derivatives and as "average bending", indicating they had developed both process and object conceptions. Others, however, lacked this understanding, often due to missing foundational ideas about second partial derivatives. This gap limited their ability to interpret the Laplacian geometrically or connect it to concavity in the temperature distribution.

Students who successfully coordinated multiple conceptions of the Laplacian—linking divergence of the gradient with second partial derivatives—formulated more coherent and physically meaningful explanations. This suggests that fostering such coordination may be a key mechanism for improving understanding of the Laplacian of the temperature.

All prerequisite mental constructions of the preliminary genetic decomposition were validated, indicating that students drew meaningfully on foundational concepts such as heat conduction, vector calculus, partial derivatives, and multivariable functions. Many elements of the heat equation PGD were also supported by student reasoning. However, students showed recurring difficulties with second partial derivatives, suggesting the need for instructional support in this area. These challenges do not indicate flaws in the PGD itself but point to opportunities for targeted scaffolding within future learning materials.

Our findings indicate that while students engage with many predicted mental constructions, certain aspects of the genetic decomposition require refinement. Specifically, current predictions for understanding heat flow (HE2) are insufficient—students often confuse zero heat flow with constant temperature. This highlights the need for a more nuanced schema of a temperature distribution (HE1) that distinguishes between steady-state, time-dependent, and equilibrium conditions. This might help students to differentiate between constant temperature and zero heat flow. Similarly, while some students progress from process to object conceptions of the temperature gradient (HE4), confusion on time-dependence suggests the need to emphasize its instantaneous nature. Finally, difficulties in activating second partial derivative schemas (Pre-iv) suggest that a more explicit focus on concavity and graphical interpretation is essential to support understanding of the Laplacian (HE7, HE8).